\documentclass[aps,10pt,prl,reprint,twocolumn,floatfix,showpacs,superscriptaddress]{revtex4-1}  

\usepackage{amsfonts,amsmath,amssymb}
\usepackage[margin=20mm,inner=20mm,marginpar=10mm]{geometry}
\usepackage{graphicx, epstopdf}
\usepackage{microtype}
\usepackage[figuresright]{rotating}
\usepackage{setspace}
\usepackage{textcomp}
\usepackage{times}
\usepackage[normalem]{ulem}
\usepackage[abs]{overpic}
\usepackage{color}
\usepackage[T1]{fontenc}
\usepackage{adjustbox}
\usepackage{longtable}
\usepackage{multirow}
\usepackage{tabularx}

\newcolumntype{L}[1]{>{\raggedright\arraybackslash}p{#1}}
\newcolumntype{C}[1]{>{\centering\arraybackslash}p{#1}}
\newcolumntype{R}[1]{>{\raggedleft\arraybackslash}p{#1}}

\newlength{\figwidth}
\begin{document}

\title{Conformer-selection by matter-wave interference}

\author{Christian Brand}
\email{brandc6@univie.ac.at}
\affiliation{University of Vienna, Faculty of Physics, Boltzmanngasse 5, A-1090 Vienna, Austria}

\author{Benjamin A. Stickler}
\email{benjamin.stickler@uni-due.de}
\affiliation{Faculty of Physics, University of Duisburg-Essen, Lotharstra\ss e 1, 47048 Duisburg, Germany}

\author{Christian Knobloch}
\affiliation{University of Vienna, Faculty of Physics, Boltzmanngasse 5, A-1090 Vienna, Austria}

\author{Armin Shayeghi}
\affiliation{University of Vienna, Faculty of Physics, Boltzmanngasse 5, A-1090 Vienna, Austria}

\author{Klaus Hornberger}
\affiliation{Faculty of Physics, University of Duisburg-Essen, Lotharstra\ss e 1, 47048 Duisburg, Germany}

\author{Markus Arndt}
\affiliation{University of Vienna, Faculty of Physics, Boltzmanngasse 5, A-1090 Vienna, Austria}
	
\date{\today}

\begin{abstract}
We establish that matter-wave diffraction at near-resonant ultraviolet optical gratings can be used to spatially separate individual conformers of complex molecules. Our calculations show that the conformational purity of the prepared partial beams can be close to 100\% and that all molecules remain in their electronic ground state. The proposed technique is independent of the dipole moment and the spin of the molecule and thus paves the way for structure-sensitive experiments with hydrocarbons and biomolecules, such as neurotransmitters and hormones, which have evaded conformer-pure isolation so far. 
\end{abstract}

\maketitle

\section{Introduction}
The conformation of a molecule can have a strong influence on its chemical reaction rates. This was demonstrated for a number of compounds~\cite{Chang_Science342_98,Taatjes_Science340_177,Sheps_PCCP16_26701,Khriachtchev_JPCA113_8143}, where the rate constants of the conformer-specific reactions varied by a factor of at least two even though the conformers differed only in the orientation of a single bond. To explore such conformer specific traits, it is desirable to develop methods to separate conformers with high efficiency from an initially unsorted molecular ensemble.

Several conformer selection techniques have been developed in recent years. Charged molecules can be separated in collision cells using ion-mobility spectroscopy~\cite{vanHelden_Science267_1483,Jarrold_PCCP9_1659,Papadopoulus_FaradayDiscuss150_243}. The separation of neutral molecules in high vacuum has been achieved using the Stark effect~\cite{Filsinger_PRL100_133003,Filsinger_AngewChemIntEd48_6900,Chang_IntRevPhysChem34_557}. This method also allows to isolate spin-isomers~\cite{Horke_AngewChemIntEd53_11965} and clusters with polar~\cite{Trippel_PRA86_033202} and non-polar~\cite{Putzke_JCP137_104310} particles in specific stochiometries. However, separation due to the Stark effect requires that the relevant conformers differ substantially in their rotational spectrum~\cite{Motsch_PRA79_013405} or their electric dipole moment~\cite{Putzke_JCP137_104310,Trippel_PRA86_033202}. Any method that can overcome these restrictions will be instrumental for subsequent slowing~\cite{Bethlem_PRL83_1558,Fulton_PRL93_243004,Hudson_PRA73_063404,Bethlem_PRL88_133003,Cherenkov_PRL112_013001,Momose_PCCP15_1772}, trapping ~\cite{vanVeldhoven_PRL94_083001,Heiner_NPhys3_115} or collision experiments~\cite{Sawyer_PCCP13_19059,Willitsch_PRL100_043203,Bell_FaradayDiscuss142_73,Kirste_Science338_1060}, for high-resolution spectroscopy~\cite{vanVeldhoven_EurPhysJD31_337,Hudson_PRL96_143004,Schnell_FaradayDiscuss150_33} and reaction studies. 

Conformer-dependent reactions of hydrocarbons and aromatic radicals can strongly affect atmospheric and astrochemical processes, such as the oxidation of aromatics~\cite{Atkinson_PolycyclAromatCompd27_15}, the formation of smog~\cite{Yu_AtmosEnviron31_2261,Bejan_PCCP5_2028}, and the chemistry on Saturn's moon Titan~\cite{Sebree_FaradayDiscuss147_231,Sebree_JACS134_1153}. Beams of size- and conformer-selected water clusters may give new insights into astrochemical~\cite{Klemperer_PNAS103_10584}, atmospheric and environmental reactions~\cite{Vaida_JCP135_020901,Wyslouzil_JCP145_211702}. Finally, small biomolecules, such as neurotransmitters and amino acids, offer rich conformational spaces~\cite{Rizzo_JCP84_2534,deVries_AnnuRevPhysChem58_585,Schwing_IntRevPhysChem35_569}. Experimental studies in the gas phase can shed new light on conformational preferences~\cite{Butz_JPCA105_544,Wilke_PCCP18_13538}, isomerization barriers~\cite{Dian_JCP120_133,Dian_Science303_1169}, and the influence of  solvent molecules in hydrated beams~\cite{Schmitt_JACS127_10356,LeGreve_JPCA113_399}.

Here, we propose to use matter-wave diffraction at a tunable standing light wave grating for separating different conformers with high purity and high efficiency. Our scheme exploits that most conformers exhibit spectroscopically well separated electronic transitions in the ultraviolet, even if they are structurally similar and have comparable dipole moments. When the laser wavelength is resonant with the electronic transition of one specific conformer, the standing light wave will realize both an absorptive and a phase grating for the matter wave~\cite{Keller_PRL79_3327,Cotter_NComms6_7336}. All other conformers will only be subjected to a pure phase grating~\cite{Gould_PRL56_827,Nairz_PRL87_160401} which can suppress certain diffraction orders depending on the laser wavelength and energy. By balancing these effects a specific diffraction order can be preferentially populated by a selected conformer. One may then select this conformer with high purity by spatially filtering the molecular interference patterns. This method can address a wide range of biomolecules, radicals, hydrocarbons, and their water clusters (Supplementary Material).

\section{Proposed Setup} 
\begin{figure}[t]
\includegraphics[width=\linewidth]{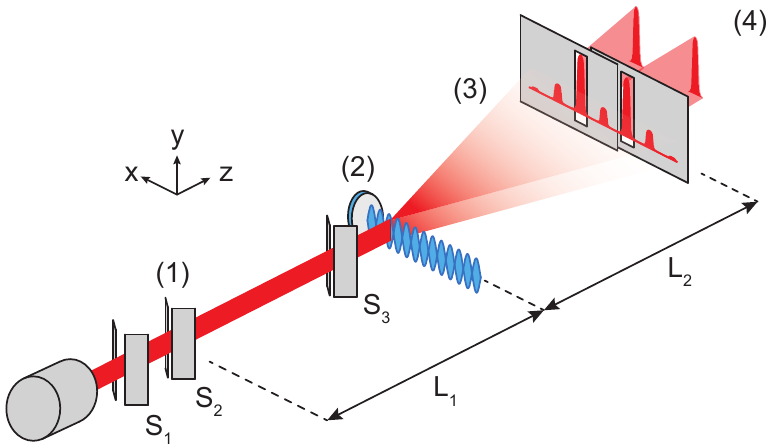}
\caption{Setup for the conformer-specific diffraction of molecules. After the first slit skimmer, the adiabatically expanded molecular beam is further position-selected by the source skimmer (1). Here, the transverse coherence of the molecular beam is prepared to illuminate several antinodes of the standing light wave (2). Matter-wave interference at the optical grating is determined by the dispersive and absorptive interaction between the molecule and the spatially periodic laser field. The resulting interference pattern (3) is spatially filtered by two movable slits to isolate the relevant diffraction peaks. This results in a pure beam of the desired conformer in the science region (4). The molecular beam is collimated by the skimmer $S_3$ to 20~$\mu$rad to prevent the diffraction orders from overlapping.}
\label{fig:Setup}
\end{figure}

In our proposed setup (see Fig.~\ref{fig:Setup}) molecules are entrained in a pulsed supersonic expansion to prepare a ro-vibrationally cold ensemble with a fast but narrow velocity distribution. Several hundred rotational levels will still be occupied due to their small energy spacing. The molecules pass a slit skimmer $S_1$ and are diffracted at the source skimmer $S_2$ in accordance with Heisenberg's uncertainty principle~\cite{Nairz_PRA65_032109}. After the distance $L_1=1$~m the molecular wave packet of mass $M$ traveling with velocity $v_z$ has a transverse coherence width $4\pi \hbar L_1/Mv_zS_2$, sufficiently large to illuminate several antinodes of a retro-reflected standing light wave. 
This optical grating with a period of $\lambda_{\rm L}/2$ can be generated by a narrow-band pulsed UV laser of wavelength $\lambda_{\rm L}$, tuned with a linewidth of better than 1~pm. State-of-the-art frequency-doubled dye lasers meet these requirements and deliver sufficient energy per pulse to diffract the molecules. At the grating the molecules are diffracted according to their de Broglie wavelength $\lambda_{\rm dB}=h/Mv$ with Planck's constant $h$. The required alignment of the grating mirror with respect to the molecular beam is determined by the pulse length of the laser and the collimation of the molecules. 
After traversing the grating, the molecular wave propagates the distance $L_2$ before it impinges onto a mask with two adjustable slits. These select, for instance, both first order diffraction peaks with a diffraction angle of $2\lambda_{\rm dB}/\lambda_{\rm L}$.

Matter-wave-diffraction at optical gratings has been realized in continuous and pulsed interferometers for atoms~\cite{Moskowitz_PRL51_370,Martin_PRL60_515}, electrons~\cite{Freimund_Nature413_142} and molecules~\cite{Haslinger_NaturePhys9_144,Nairz_PRL87_160401,Gerlich_NPhys3_711}. The phase grating transfers an integer multiple of the grating momentum $2 h/\lambda_{\rm L}$ onto the traversing, rotating molecule. In addition, the molecule may absorb one or more photons depending on the laser wavelength, intensity, molecular absorption cross section, and interaction time. If only a single photon is absorbed, the coherent diffraction signal will be shifted in momentum by $\pm h / \lambda_{\rm L}$ and filtered out by the mask provided that the diffraction peaks are sufficiently separated. Upon absorption of two or more photons, the molecule is ionized and removed from the beam.

\section{Molecule-Laser Interaction} The force acting on a polarizable molecule in the laser field is determined by the optical susceptibility, which depends on the laser wavelength $\lambda_{\rm L}$ and the rotational state $r$. The real part of the susceptibility is the polarizability $\alpha_r(\lambda_{\rm L})$ while the imaginary part $\varepsilon_0 \lambda_{\rm L} \sigma_r(\lambda_{\rm L})/2 \pi$ depends on the total absorption cross section $\sigma_r(\lambda_{\rm L})$.

The effect of the pulsed standing light wave on the transverse motional state of the molecule can be described by the phase shift $\phi_r(\lambda_{\rm L},E_{\rm L})$ and the mean number of absorbed photons $n_r(\lambda_{\rm L},E_{\rm L})$ at the antinodes as a function of $\lambda_{\rm L}$ and the pulse-energy $E_{\rm L}$. The grating transit of a molecule in a rotational state $r$ is then characterized by the state-dependent grating transformation~\cite{Nimmrichter_NJPhys13_075002,Stickler_PRA92_023619}
\begin{eqnarray}
t_r(x)&=&\exp\left[\left(i\phi_r-\frac{n_r}{2}\right)\cos^2\left(\frac{2\pi x}{\lambda_{\rm L}}\right)\right].
\end{eqnarray}
The phase and the mean photon number can be related to the real and imaginary part of the susceptibility,
\begin{equation}
\phi_r(x,\lambda_{\rm L},E_{\rm L})=\frac{2 \alpha_r(\lambda_{\rm L})E_{\rm L}}{\hbar\epsilon_0 c A_{\rm L}},
\end{equation}
and
\begin{equation}
n_r(\lambda_{\rm L},E_{\rm L})=\frac{2\sigma_r(\lambda_{\rm L})E_{\rm L}\lambda_{\rm L}}{\pi\hbar c A_{\rm L}},
\end{equation}	
with the spot size $A_{\rm L}$.

\section{Interference Pattern} The transverse wavefunction of a molecule's center-of-mass motion, starting at position $x_0$ with velocity $v_z$ in the rotational state $r$ can be evaluated in the paraxial approximation as~\cite{Hornberger_PRA70_053608,Stickler_PRA92_023619}
\begin{eqnarray}
\Psi_r(x;x_0)&\propto &\int_{-S_3/2}^{S_3/2}{\!\!\!\!dx^\prime t_r(x^\prime)}\exp\left[\frac{iMv_z}{2\hbar}\left(\frac{1}{L_1}+\frac{1}{L_2}\right)x^{\prime 2}\right] \notag \\
&&\times \exp \left [- \frac{i M v_z }{\hbar}\left(\frac{x_0}{L_1}+\frac{x}{L_2}\right)x^\prime\right ].
\end{eqnarray}
Here, $S_3$ is the total width of the grating as determined by the third skimmer. Averaging over all source points and rotation states with population $p_r$ then yields the molecular interference pattern
\begin{equation}
S(x) \propto \int_{-S_2/2}^{S_2/2}\text{d}x_0 \sum_r p_r \vert\Psi_r(x;x_0)\vert^2.
\end{equation}
 
Interferometric conformer selection can be applied to all molecules with spectroscopically separated transitions in the visible/ultraviolet. This can be seen by considering the approximate intensity $S_n$ of the $n$-th diffraction order in the far-field, {\it i.e.} for $M v_z(L_1 + L_2)S_3^2/2 \hbar L_1 L_2 \ll 1$,
\begin{equation}\label{eq:farfield}
 S_n(\lambda_{\rm L}, E_{\rm L}) \propto \sum_r p_r e^{- n_r/2} \left \vert J_n \left ( \frac{ \phi_r}{2} + i \frac{n_r}{4} \right ) \right \vert^2,
\end{equation}
where $J_n(\cdot)$ are Bessel functions. Spectroscopic separability of the conformers implies that the wavelength $\lambda_{\rm L}$ can be tuned in such a way that the polarizability of any chosen conformer exceeds all others substantially. By 
adjusting the pulse energy sufficiently low so that the Bessel function in Eq.~\eqref{eq:farfield} vanishes for all but the selected species, the first diffraction order will be populated predominantly by this conformer. This demonstrates the general applicability of the proposed method, even though there can be regions of $\lambda_{\rm L}$ and $E_{\rm L}$ for which interferometric conformer selection works with even higher efficiency.

\section{Conformer-selective Interference of PEA} 
\begin{figure}[bt]
\includegraphics[width=\linewidth]{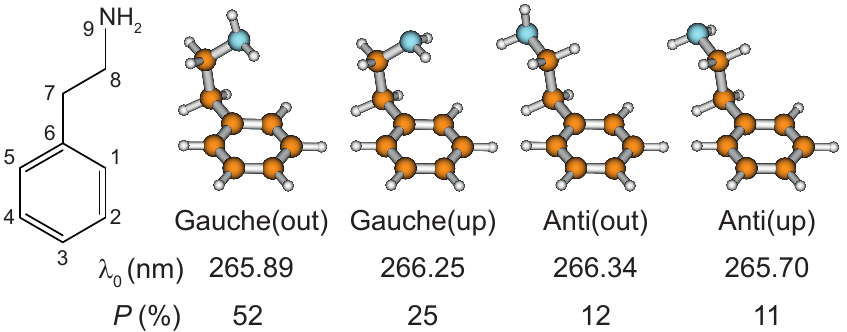}
\caption{Conformer selection by matter-wave interference is illustrated for the example of 2-phenylethylamine (PEA). In a supersonic expansion it reveals four spectroscopically separated conformers, which differ in the orientation of the C--N bond with respect to the C$_6$--C$_7$ bond (atomic numbering on the left) and in the lone pair of the amino group relative to the chromophore. All conformers can be isolated in matter-wave diffraction, even though their  dipole moments are virtually identical. Below each conformer we note the electronic transition energy $\lambda_0$ and the relative population $P$~\cite{Dickinson_JACS120_2622} in molecular beams.}
\label{fig:conformers}
\end{figure}

To illustrate the method, we study the neurotransmitter 2-phenylethylamine (PEA), see Fig.~\ref{fig:conformers}. In jet experiments it exhibits four conformers, which are spectroscopically well separated~\cite{Martinez_JMolSpectrosc158_82,Dickinson_JACS120_2622}. They differ in the conformation of the C--N bond and the lone pair of the NH$_2$ group. While the former is either in \emph{anti} or \emph{gauche} position to the C$_6$--C$_7$ bond, the latter points \emph{up} or \emph{out} with respect to the chromophore. Since PEA has only a single polar group, the dipole moments of all four observed conformers are predicted to be virtually identical ($1.25\pm 0.05$~D)~\cite{Lopez_PCCP9_4521,Brand_JPCA115_9612}. Stark separation is thus practically impossible~\cite{Putzke_JCP137_104310}. The susceptibility of the four different conformers is calculated from data compiled in the Supplementary Material. 
The complex optical susceptibility of each conformer can be described by a Lorentz oscillator model~\cite{Townes1975,Bohren2007} where the transition frequencies of the rotational states are calculated from the conformer-dependent rotational constants~\cite{meerts2006,Dickinson_JACS120_2622}. The mean static polarizability $\alpha_0/4\pi\epsilon_0 = 14.6$~\AA$^3$ was calculated using density functional theory and agrees well with the value obtained from the refractive index in solution~\cite{CRCHandbook97}. The maximum deviation between different conformers is at $0.2$~\AA$^3$ (Supplementary Material) and negligible compared to the optical polarizability which changes by tens of \AA$^3$ over the considered wavelength region. The rotational sub-structure of the electronic transition consists of several hundred individual peaks even at a rotational temperature of $3$~K. To account for this, we divide the rotational spectrum into 20 intervals and calculate for each bin its mean spectroscopic weight $p_{\rm r}$.

\section{Conformer Selection Efficiency} 
\begin{figure*}[t]
\includegraphics[width=\linewidth]{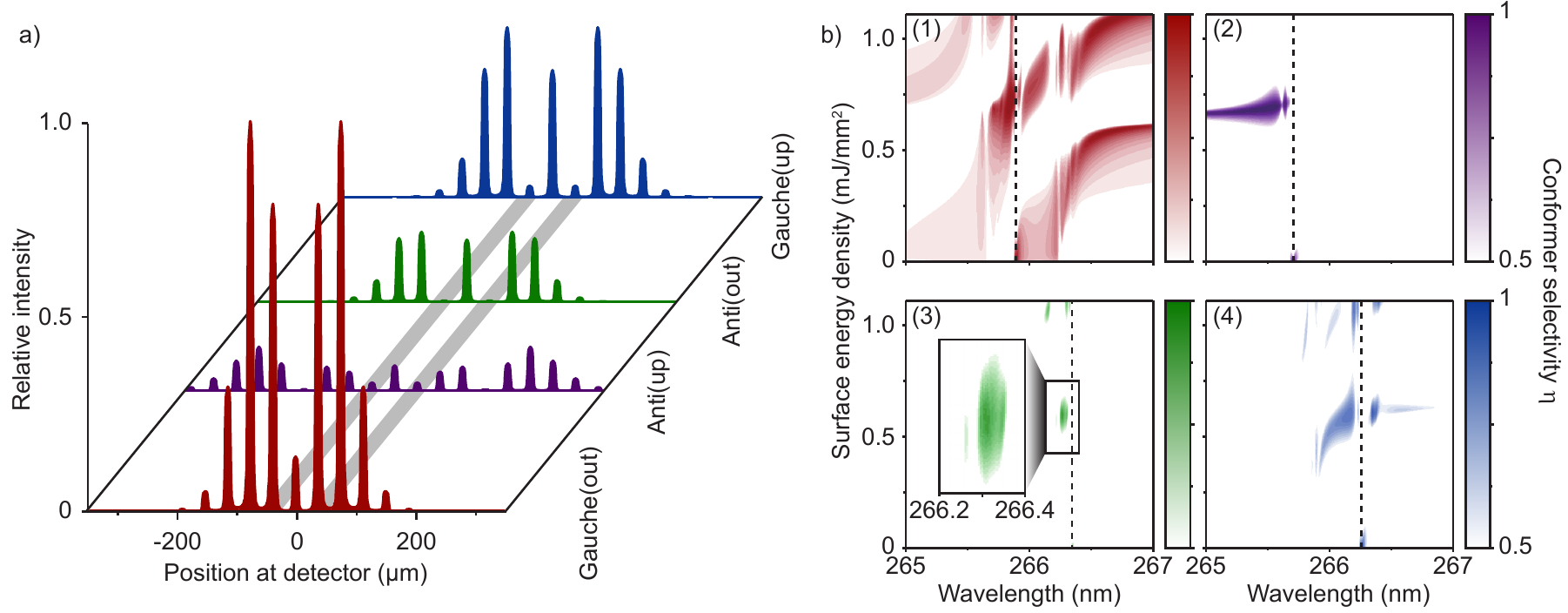}
\caption{a) Molecular interference patterns of the four conformers of PEA for a surface energy density $E_{\rm L}/A_{\rm L}$ of $0.67$~mJ/mm$^2$ at $\lambda_{\rm L}=265.8$~nm. Here the first diffraction orders is preferentially populated by the Gauche(out) conformer (red) with a conformer-selection efficiency $\eta$ of $>93\%$ (grey stripes). b) Conformer-selection efficiency $\eta$ in the first diffraction order for the (1) Gauche(out), (2) Anti(up), (3) Anti(out), and (4) Gauche(up) conformer. Areas with a selectivity of less than 50\% are left blank. The position of the electronic resonance is indicated by the broken line. For all conformers a selectivity of a least 80\% and up to 99\% is predicted, corresponding to an up to 9-fold increase in the relative population. The respective areas in parameter space are easily accessible in an experiment even for conformers of low abundance, as shown in the inset for Anti(out).}
\label{fig:diffraction_all}
\end{figure*}

In Fig.~\ref{fig:diffraction_all}a we show the interference patterns of the four conformers diffracted at a grating with $\lambda_{\rm L}=265.8$~nm, $E_{\rm L}=0.67$~mJ, and $A_{\rm L}= 1$~mm$^2$. The grating wavelength lies between the electronic transitions of the Anti(up) and the Gauche(out) conformers. Hence, the respective values of $\alpha_r(\lambda_{\rm L})$ deviate strongly from the static polarizability, while they are close to that value for Gauche(up) and Anti(out). At the chosen parameters the intensity in the first diffraction order is maximal for Gauche(out) comprising 35\% of the total population of this conformer.

From the interference patterns we calculated the \emph{conformer-selection efficiency} $\eta$ in the first diffraction order, that is the fraction of one conformer compared to the sum of all four. For the patterns shown in Fig.~\ref{fig:diffraction_all}a $\eta$ reaches a value of $>93\%$, illustrating that high-purity conformer-selection is possible with this method. To prove the feasibility of a clear separation for all molecular conformers, we have extended the simulations to the wavelength region between 265 and 267~nm and to an surface energy density $E_{\rm L}/A_{\rm L}$ up to 1.1~mJ/mm$^2$. For each simulated pattern we calculated $\eta$ 
and compiled the results in Fig.~\ref{fig:diffraction_all}b for the four conformers. Here we color the plot only when an efficiency of 50~\% is exceeded, areas of lower selectivity are left blank. It shows a rich pattern due to the strong wavelength-dependence of $\sigma_{\rm r}(\lambda_{\rm L})$ and $\alpha_{\rm r}(\lambda_{\rm L})$ near the resonances. The size of the parameter space which leads to selectivity depends strongly on the relative population of the conformer in the beam. For the Gauche(out) conformer a selectivity greater than 50\% is observed in large parts of the parameter space. However, even for the weakly populated Anti(out) conformer $\eta$ exceeds 80\% over a range of 0.03~nm which is experimentally easily accessible. It demonstrates that every single conformer of PEA can be selected with a high conformer selectivity and that specific conformers can be addressed by tuning the laser wavelength and power. Since the details of the selection efficiency shown in Fig.~\ref{fig:diffraction_all}b are highly sensitive on the oscillator strength $f$, measuring the conformer-selection efficiency $\eta$ offers a new way to probe it for all conformers.

\section{Experimental Feasibility} The separation between neighboring resonances of the polarizability spectrum is on the order of $0.1$~nm, and therefore well resolved by a dye laser with a laser linewidth of 1~pm, even in the presence of rotational broadening of the spectrum. For medium-sized molecules such as PEA, the rotational energy spread at 3~K is roughly 0.02~nm~\cite{Dickinson_JACS120_2622} -- much smaller than the separation between the electronic resonances of the neighboring conformers.  
In a realistic experimental situation a certain fraction of the molecular beam will not interact with the pulsed laser beam and pass on to the detector in the blocked zeroth diffraction order. Furthermore, as the carrier gas is much lighter than the molecules, it experiences larger diffraction angles than the analyte molecule. Hence, the diffraction orders do not overlap and the proposed method is virtually background-free. This is comparable to Stark deflection where polar conformers are deflected out of the initial beam~\cite{Filsinger_AngewChemIntEd48_6900}. 

The selectivity of our proposed method depends on the power stability of the laser system. When operated in saturation the fluctuation of a dye laser depends solely on the performance of the diode-pumped solid state laser, which can reach $<3\%$. After frequency doubling this leads to $\pm 6\%$ stability in power. However, even for power variations of $\pm 15\%$ the simulations predict values of $\eta$ between $70\%$ for Anti(out) and $95\%$ for Gauche(out) (see Supplementary Material). 

The flux for a specific conformer behind the selection slits can be estimated using the vapor pressure of PEA at 410~K~\cite{Mokbel_JChemEngData54_819}, the characteristics of a pulsed Even-Lavie-valve~\cite{Hillenkamp_JCP118_8699,Even_EPJTechInstrum2_17}, a 100~Hz laser illuminating an area of $1$~mm$^2$, and the diffraction pattern of the Gauche(out) conformer shown in Fig.~\ref{fig:diffraction_all}a. Assuming that 78\% of the molecules are in their vibrational ground state and a mean number of absorbed photons of $< 10^{-4}$, we expect a mean flux of $1.3\times 10^{9}$~cm$^{-2}$s$^{-1}$ (Supplementary Material). The peak density reaches $1.3\times 10^{8}$~cm$^{-3}$ which is comparable to densities used in X-ray diffraction at free electron lasers~\cite{Kupper_PRL112_083002} and in crossed beam studies \cite{Kirste_Science338_1060}. 

Since the presented matter-wave assisted separation technique requires the conformers only to have sharp and separated transitions in the ultraviolet, it can be used for a wide range of species and applications. This includes large families of molecular systems such as hydrocarbons, small biomolecules, and aromatic radicals~\cite{deVries_AnnuRevPhysChem58_585,Zwier_JPCA110_4133,Sebree_ChemSci2_1746} (Supplementary Material). It can provide conformer-pure samples for X-ray diffraction~\cite{Barty_AnnuRevPhysChem64_415} or Coulomb explosion studies~\cite{Levin_PRL81_3347,Pitzer_Science341_1096}, merged beam experiments~\cite{Kirste_Science338_1060,Wei_JCP137_054202} or collisions with trapped neutral particles~\cite{Sawyer_PCCP13_19059} or cold ions~\cite{Chang_Science342_98,Willitsch_PRL100_043203,Bell_FaradayDiscuss142_73}. 
The selected structure can be varied during the experiment by changing the wavelength and the pulse energy of the grating laser. This way several conformers can be compared within one experimental run. The combination with molecular cooling schemes might also prepare samples for high-resolution spectroscopy~\cite{vanVeldhoven_EurPhysJD31_337,Hudson_PRL96_143004,Schnell_FaradayDiscuss150_33}.

\section{Conclusion} We have presented a robust scheme to select molecular conformers in the vibrational ground state with purities of up to 95\%. The method requires an intense and tightly collimated molecular beam giving rise to matter-wave diffraction at a tunable laser grating. It is applicable to large families of conformers that can be distinguished by individual and sharp electronic transitions in the UV. Being independent of internal dipole moments or the spin state, it can be applied to non-polar molecules, radicals and their clusters alike. Furthermore, this technique eliminates most of the vibrationally excited molecules from the region of interest, selecting colder molecules in the beam. 

\section{acknowledgments}

C.B. and B.A.S. contributed equally to this work. This project has received funding from the European Research Council (ERC) under the EU Horizon 2020 research and innovation programme (grant agreement n$^\circ$320694) and the Austrian Science Funds FWF W1210-N25. 


%

\newpage
\section{Supplementary Material}
\subsection{Optical polarizability and rotational average for 2-phenylethylamine (PEA)}

For each conformer of PEA, the complex optical polarizability at laser wavelength $\lambda_{\rm L} =  2 \pi c/ \omega_{\rm L}$ is calculated using the Lorentz-Drude formula for a single electronic transition~\cite{CRCHandbook97,Townes1975,Bohren2007},
\begin{equation} 
\alpha(\omega_{\rm L}) = \alpha_0 + \frac{e^2_0 f}{m_e} \frac{1}{\omega^2_0 - \omega^2_{\rm L} - i \omega_{\rm L}/ \tau}.
\label{eq:optpol}
\end{equation}
We assumed a common excited state lifetime $\tau$ of 70~ns based on experimental data~\cite{Martinez_JMolSpectrosc158_82}. The static polarizabilities were calculated using density functional theory with the PBE0 functional employing the Def2-TZVP basis set and are compiled in Table~\ref{tab:data}. The oscillator strength $f$ and the electronic transition wavelength $\lambda_0 =  2 \pi c/ \omega_{\rm 0}$ are also compiled in Table~\ref{tab:data} for each conformer of PEA. 

In order to take the rotational structure of the transition from the electronic ground to the excited state $(g, r) \to (e,r')$ into account, we note that the final measurement traces over all rotation states. Thus, any coherences between rotational levels can be neglected and we determine the interference pattern for each transition, replacing $\omega_0$ by $\omega_{rr'} = (E_{er'} - E_{gr})/\hbar$ in \eqref{eq:optpol}, and sum over the resulting interference patterns, weighted by their respective spectroscopic relative intensity. Several hundred transitions relevant for PEA are collected in a histogram. We use the mean frequency and spectroscopic weight of each bin to perform the calculation. Only in the case of rotation-state resolved detection, the coherences between rotational levels need to be considered.

\begin{table}[b]
\caption{Calculated oscillator strength $f$~\cite{Brand_JPCA115_9612}, electronic transition energy $\lambda_0$, static polarizability $\alpha_0/(4\pi\epsilon_0)$, and relative population $P$~\cite{Dickinson_JACS120_2622} in molecular beams of the four experimentally observed~\cite{Martinez_JMolSpectrosc158_82,Dickinson_JACS120_2622} conformers of PEA.}
\begin{tabular}{l|cccc}
Conformer 	& $f\times 10^3$	&	$\lambda_{0}$ (nm)	&	$\alpha_0$ (\AA$^3$)	&	$P (\%)$ 	\\
\hline
Gauche(out)	&	0.56						&	265.89							&	14.5									& 52	\\
Gauche(up)	&	0.92						&	266.25							&	14.5									& 25	\\
Anti(out)		&	0.64						&	266.34							&	14.6									& 12	\\
Anti(up)		&	1.94						&	265.70							&	14.7									& 11	
\end{tabular}
\label{tab:data}
\end{table}

\subsection{Selection efficiency}
\begin{figure}
\includegraphics[width=0.91\linewidth]{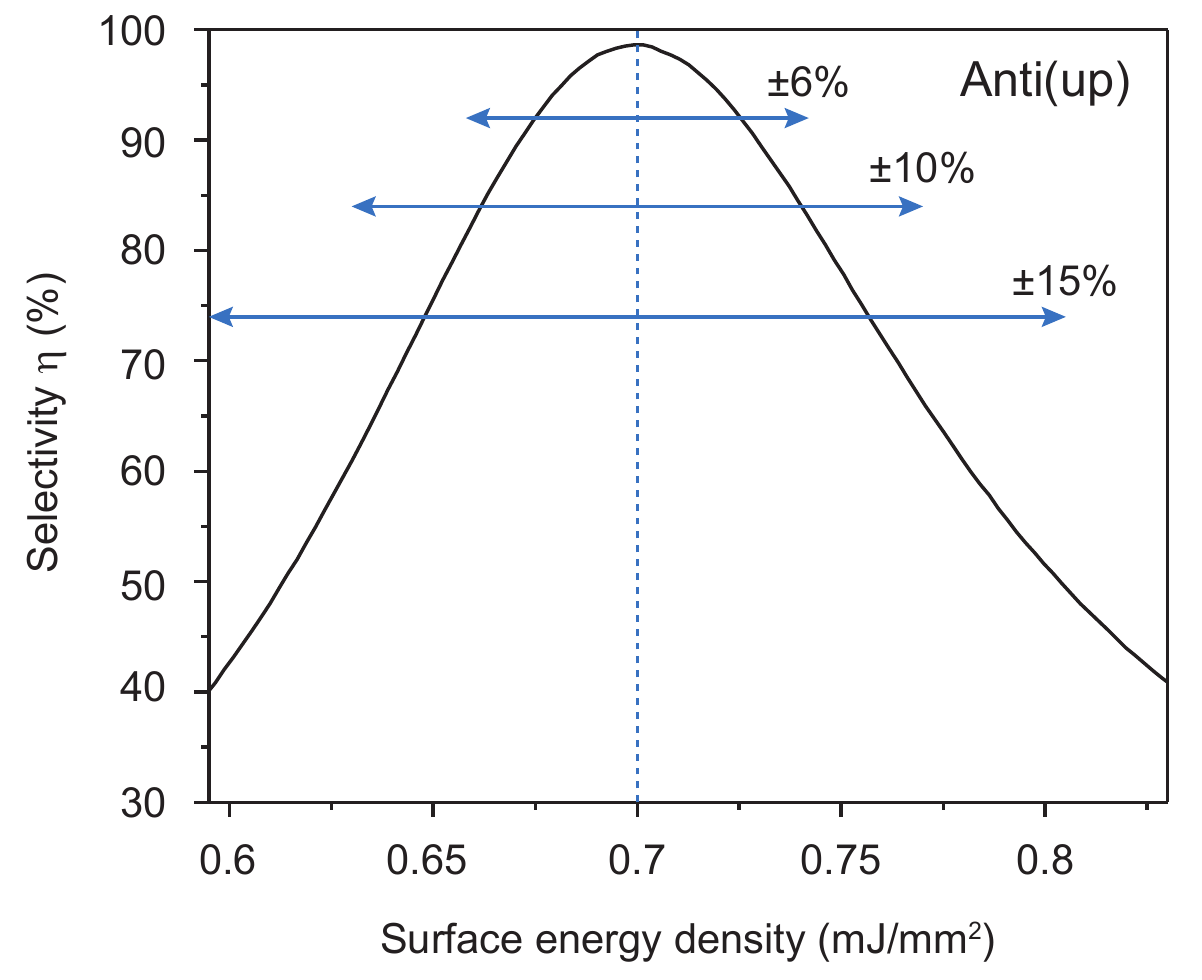}
\caption{Mean selectivity $\eta$ for the Anti(out) conformer at a laser wavelength of $\lambda_{\rm L}=265.55$~nm depending on the surface energy density. At 0.7~mJ/mm$^2$ (dashed line) a maximum selectivity of 99\% is reached. The integrated selectivity due to power variations between 6 and 15\% is indicated by horizontal lines. It ranges between 92 and 74\%.}
\label{fig:Selectivity}
\end{figure}
The selectivity of the proposed method depends on both the wavelength $\lambda_{\rm L}$ of the grating laser and the surface energy density. The wavelength stability of current pulsed dye lasers is in the pm regime which is sufficient to maintain stable diffraction pattern over time. The effect of power fluctuations is illustrated in Fig.~\ref{fig:Selectivity} for the Anti(out) conformer at a wavelength of 265.55~nm. The selectivity peaks at a surface density of 0.7~mJ/mm$^2$. Power fluctuations of $\pm 6\%$ would reduce the integrated selectivity to 92$\%$. Even under strong energy fluctuations, for instance, at $0.7$~mJ/mm$^2\pm15\%$ the selectivity still reaches $74\%$. The power-dependent selectivities of all four conformers are compiled in Table~\ref{tab:selection-efficiency}.     
\begin{table}[ht]
\caption{Influence of power variations on the selection efficiency $\eta$ for PEA. For each conformer the selectivity is stated at the center surface power density and averaging over region of $\pm 10~\%$ and $\pm 15~\%$.}
\begin{tabular}{l|cllcc}
\hline\hline
Conformer 	& Wavelength	&&	Surface power 			&&	Mean  				\\
						& (nm)				&&	density (mJ/mm$^2$)	&& selectivity 	\\
\hline                                            
Gauche(out)	& 265.84			&&	$0.72$						  && 100 \\
						&							&&	$0.72\pm10\%$				&& 97	\\
						& 						&&	$0.72\pm15\%$				&& 95	\\ [0.3em]
Gauche(up)	& 266.17			&&	$0.63$							&& 83	\\
						& 						&&	$0.63\pm10\%$				&& 79	\\
						& 						&&	$0.63\pm15\%$				&& 75	\\ [0.3em]
Anti(out)		& 266.26			&&	$0.61$							&& 88	\\
						& 						&&	$0.61\pm10\%$				&& 77	\\
						& 						&&	$0.61\pm15\%$				&& 70	\\ [0.3em]
Anti(up)		& 265.55			&&	$0.70$							&& 99	\\	
						& 						&&	$0.70\pm10\%$				&& 84	\\
						& 						&&	$0.70\pm15\%$				&& 74	\\
\hline\hline
\end{tabular}
\label{tab:selection-efficiency}
\end{table}

\subsection{Expected molecular flux behind the selection slits}

The particle flux $j=nv=pv/k_{\rm B}T$ is determined by the partial pressure $p$ of 2-phenylethylamine, the temperature $T$, and the velocity $v$ of the molecules. Taking the velocity of an adiabatic expansion of Argon ($v=650$~m/s) and the partial pressure of 2-phenylethylamine at $T=410$~K \cite{Mokbel_JChemEngData54_819}, this yields a flux of $j = 2.4\times 10^{27}$~m${}^{-2}$s${}^{-1}$.

We consider the example of a supersonic expansion from an Even-Lavie valve~\cite{Even_EPJTechInstrum2_17} with a nozzle diameter of $50~\mu$m, emitting a molecular current of $4.6\times 10^{18}$~s${}^{-1}$. The spatial distribution of the expansion can be described by a gaussian with a full width at half maximum of 12$^\circ$~\cite{Hillenkamp_JCP118_8699}. At the second collimation slit S$_2$, 0.1~m behind the nozzle, the molecular beam has a half width at half maximum of 10~mm. The fraction of molecules transmitted through the slit of 1~mm height and $10~\mu$m width is calculated from the two-dimensional integral over the size of the detector slit.
\begin{figure}[htb]
\includegraphics[width=0.9\linewidth]{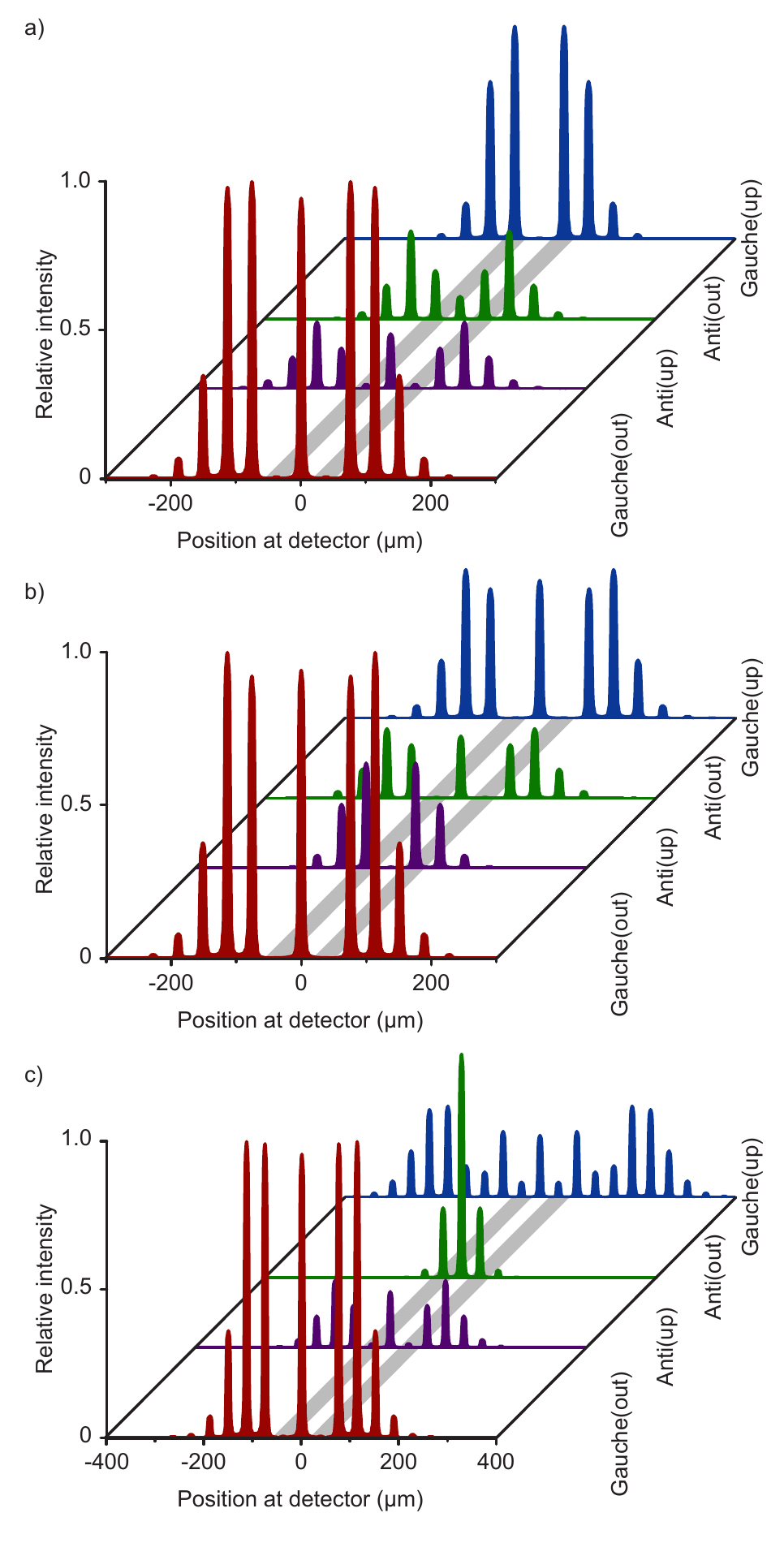}
\caption{Molecular diffraction patterns of the four conformers of 2-phenylethylamine at parameters of high selectivity for one specific conformer according to Fig.~\ref{fig:diffraction_all}b. At $\lambda_{\rm L}=266.1$~nm and 0.57~mJ/mm$^2$ the Gauche(up) conformer is selected with high selectivity (a). The diffraction patterns in (b) and (c) show the points of high selectivity for the Anti(up) conformer ($\lambda_{\rm L}=265.5$~nm and 0.7~mJ/mm$^2$) and the Anti(out) conformer ($\lambda_{\rm L}=266.3$~nm and 0.61~mJ/mm$^2)$.}
\label{fig:Beugungsbilder_remaining}
\end{figure}
This yields a reduction factor of $2.2\times 10^{-5}$, leading to a particle current $I = 1.0\times 10^{14}$ s${}^{-1}$ for a continuous flow of molecules. Here we consider
that the skimmer S$_1$ does not play a significant role as S$_2$ leads to a more stringent collimation. The collimated beam propagates 1~m until it reaches the skimmer S$_3$ (1~mm $\times$ 10~$\mu$m), where the number of particles is reduced by another factor of $1.4\times10^{-3}$. This yields an overall reduction factor of $3.0\times10^{-8}$. For a spot size of $\ell^2 = 1\times 1$~mm$^2$, the number of molecules illuminated by the laser is $I \ell/v = 215\,500$ molecules per pulse. The number of isolated molecules also depends on the relative population of the targeted conformer, the population of the vibrational ground state, the number of molecules ionized by the grating laser, and the percentage of molecules diffracted into the first order.

In Fig.~\ref{fig:diffraction_all}a we consider the diffraction of the Gauche(out) conformer of PEA. Its relative population is 52\%~\cite{Dickinson_JACS120_2622}, 78\% of all molecules are expected to be in the vibrational ground state, and 35\% of the conformers are diffracted into the first order. The ionization probability at this wavelength is computed to be below $10^{-4}$ and can be neglected. This yields about $30\,800$ molecules per laser shot and a mean current of $1.3\times 10^{9}$ s${}^{-1}$~cm${}^{-2}$, assuming a laser repetition rate of 100~Hz and a detection area of $2$~mm $\times$ $115$~$\mu$m. The width of the area is given by separation of the first diffraction orders. The peak density corresponds to $1.3\times 10^{8}$ cm$^{-3}$ which compares well to the beam density used in XUV diffraction experiments at free electron lasers ($8.0\times 10^{7}$ cm$^{-3}$)~\cite{Kupper_PRL112_083002}. 
As shown in Fig.~\ref{fig:diffraction_all}b, also less abundant conformers can be isolated with high selectivity. Fig.~\ref{fig:Beugungsbilder_remaining}a shows the molecular diffraction patterns for all conformers at $\lambda_{\rm L}=266.1$~nm and a surface power density of 0.57~mJ/mm$^2$. At these parameters the Gauche(up) conformer is selected and 51\% of all molecules in the Gauche(up) conformation are in the first diffraction orders, leading to a peak density of $9.4\times 10^{7}$ cm$^{-3}$. The diffraction patterns at points of high selectivity for the two Anti conformers are shown in the lower panels. The extracted peak density is $4.6\times 10^{7}$ cm$^{-3}$ for Anti(up) and $3.3\times 10^{7}$ cm$^{-3}$ for Anti(out).

\begin{figure}
\includegraphics[width=\linewidth]{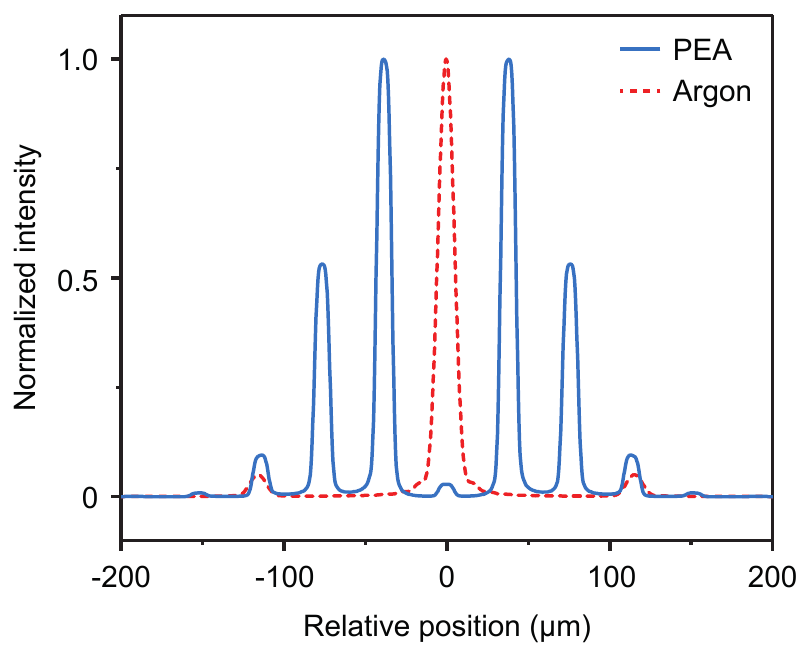}
\caption{Comparison of the diffraction pattern of 2-Phenylethylamine (PEA) and Argon at a phase grating with $\lambda_{\rm L}=266$~nm. The first diffraction order of Ar overlaps with the third of PEA due to their mismatch in mass.}
\label{fig:Vergleich}
\end{figure}

In Fig.~\ref{fig:Vergleich} the diffraction patterns of PEA and Argon at a phase grating with period 266~nm are compared. As the spacing of the diffraction orders scales inversely with mass, the backing gas is diffracted to larger angles. Hence, the diffraction orders do not overlap, preventing the gas from reaching the detection area.

\subsection{Suitable molecular systems}

In Table~\ref{tab:systems_1}-\ref{tab:systems_3} we list a number of molecules which can be addressed with the proposed selection method. It contains a set of systems systems which are of importance for several fields of physics and chemistry and is by no means exhaustive. The chosen particles molecules have sufficiently high vapor pressure and are light enough to be compatible with fast molecular beams. We have compiled all conformers even though not all of them might be isolated, either because their relative population is too small or they are spectrally overlapping with other conformers. This list can easily be extended to clusters with noble gas atoms, and small molecules like O$_2$, N$_2$, and CH$_4$. Often these can be prepared in high abundance by co-expanding them with the desired molecule. As the clusters do not absorb a photon and remain in their electronic ground state during the diffraction, fragmentation can be neglected. 

\begin{table*}
\caption{Molecular systems and their water clusters suitable for conformer selection.}
\label{tab:systems_1}
\begin{tabular}{L{4cm}C{5cm}cC{4cm}C{3cm}}
\hline\hline
Substance						&	Structure																																&	&																																		& Electronic origin (cm$^{-1}$)	\\
\hline
\multicolumn{5}{c}{\underline{Neurotransmitter and hormones}}\\  [0.5em]
2-Phenylethylamine	&	\multirow{4}{*}{\includegraphics[width=3.063cm]{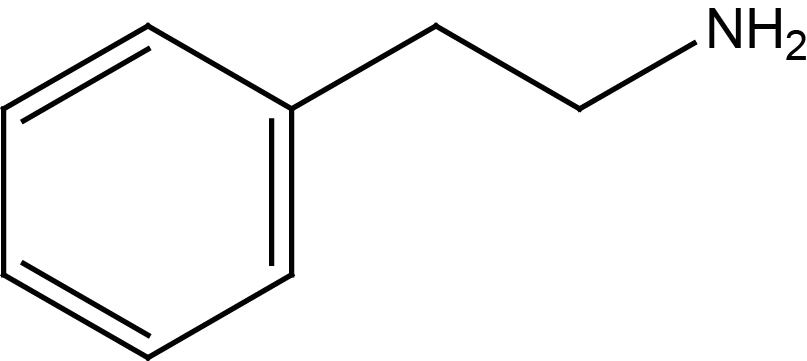}}				&	&	4 conformers~\cite{Sipior_JFluorescence1_41,Dickinson_JACS120_2622} &	37\,546				\\
										&																																					&	&																																			& 37\,558				\\ 
										&																																					&	&																																			& 37\,610				\\ 
										&																																					&	&																																			& 37\,636				\\ 	[0.3em]
										&																																					&	&	(H$_2$O)$_1$-cluster~\cite{Dickinson_JACS120_2622}									& 37\,630				\\	[0.3em]
Serotonin						& \multirow{4}{*}{\includegraphics[width=4.07cm]{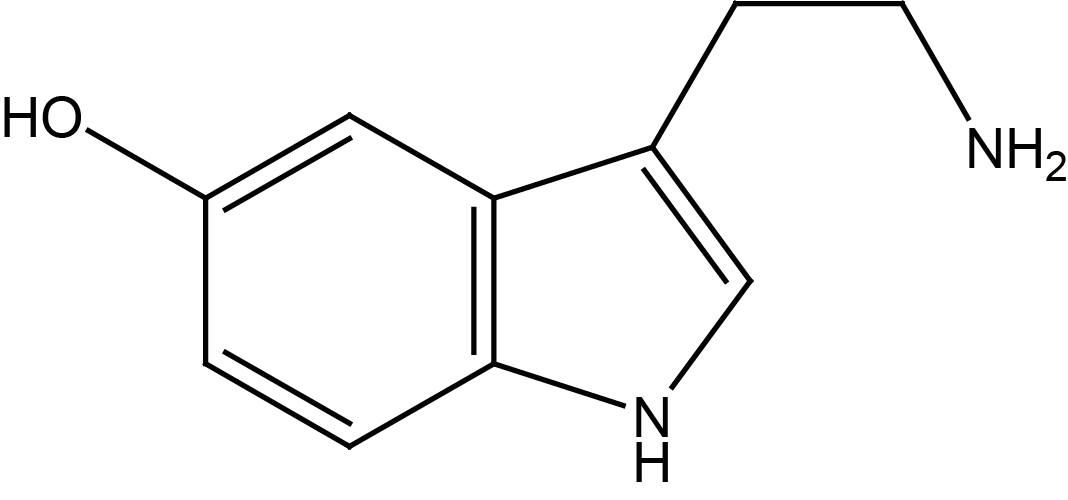}}		&	&	8 conformers~\cite{LeGreve_JACS129_4028} 														&	32\,282				\\
										&																																					&	&																																			& 32\,313				\\ 
										&																																					&	&																																			& 32\,353				\\ 
										&																																					&	&																																			& 32\,537				\\ 
										&																																					&	&																																			& 32\,545				\\ 
										&																																					&	&																																			& 32\,548				\\ 
										&																																					&	&																																			& 32\,560				\\ 
										&																																					&	&																																			& 32\,584				\\ [0.3em]
										&																																					&	& (H$_2$O)$_1$-cluster~\cite{LeGreve_JPCA113_399}											& 32\,183				\\
										&																																					&	&																																			& 32\,449				\\ 
										&																																					&	&																																			& 32\,666				\\ [0.3em]			Melatonin						&	\multirow{4}{*}{\includegraphics[width=5.102cm]{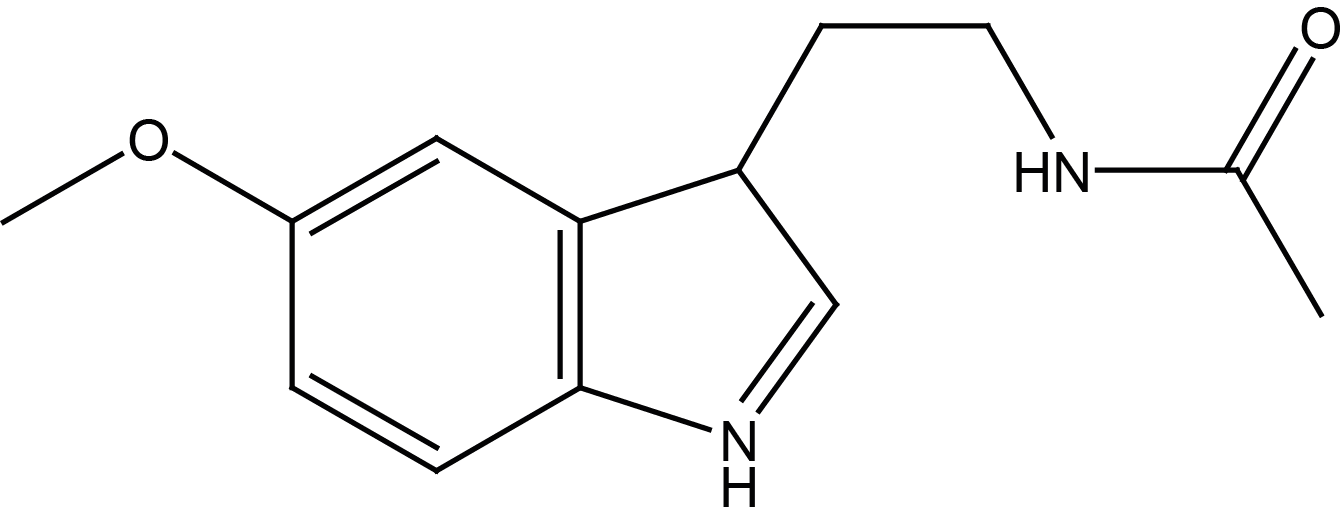}}	&	& 5 conformers~\cite{Florio_JACS124_10236} 														&	32\,432				\\
										&																																					&	&																																			& 32\,483				\\ 
										&																																					&	&																																			& 32\,614				\\ 
										&																																					&	&																																			& 32\,621				\\ 
										&																																					&	&																																			& 32\,795				\\ [0.3em]
										&																																					&	&	(H$_2$O)$_1$-cluster~\cite{Florio_JPCA107_974}											& 32\,442				\\ 
										&																																					&	&																																			& 32\,673				\\ 
										&																																					&	&																																			& 32\,842				\\ 
										&																																					&	&																																			& 32\,956				\\ [0.3em]	
Tryptamine					&	\multirow{4}{*}{\includegraphics[width=3.281cm]{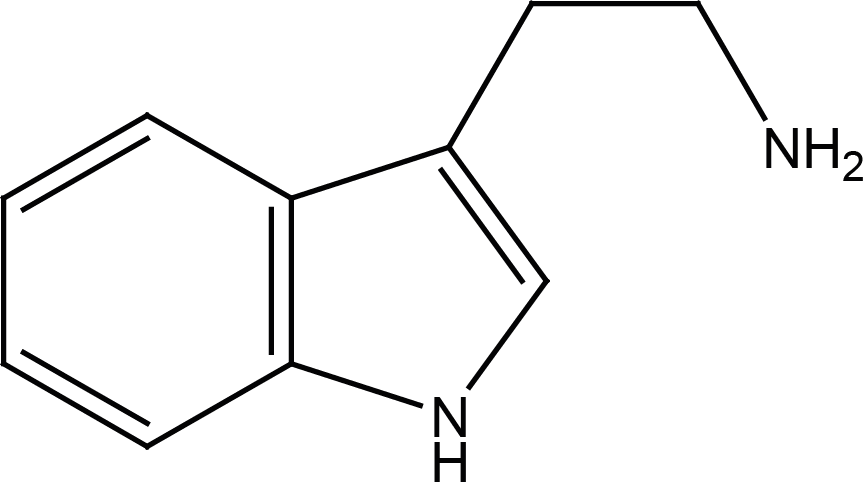}}	&	&	 7 conformers~\cite{Park_JCP84_6539,Carney_CPL341_77,Nguyen_MolPhys103_1603}	&	34\,832				\\ 
										&																																					&	&																																			& 34\,868				\\ 
										&																																					&	&																																			& 34\,879				\\ 
										&																																					&	&																																			& 34\,880				\\ 
										&																																					&	&																																			& 34\,884				\\
										&																																					&	&																																			& 34\,896				\\ 
										&																																					&	&																																			& 34\,916				\\ [0.3em]
										&																																					&	&	(H$_2$O)$_1$-cluster~\cite{Sipior_JCP88_6146}												& 34\,957				\\ [0.3em]	
Mexamin							&	\multirow{3}{*}{\includegraphics[width=4.385cm]{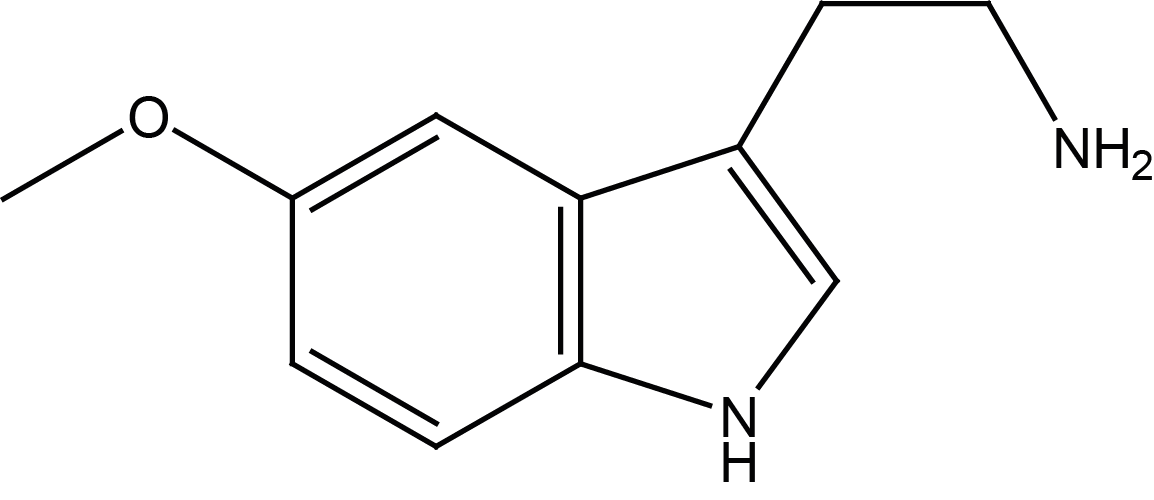}}		&	& 3 conformers~\cite{Vu_PCCP11_2433} 																	&	32\,734				\\ 
										&																																					&	&																																			& 32\,764				\\
										&																																					&	&																																			& 32\,808				\\ [0.3em]
										&																																					&	&	(H$_2$O)$_1$-cluster~\cite{Vu_PCCP11_2433}													& 32\,528				\\ 
										&																																					&	&																																			& 							\\ 
Amphetamine 				&	\multirow{3}{*}{\includegraphics[width=3.063cm]{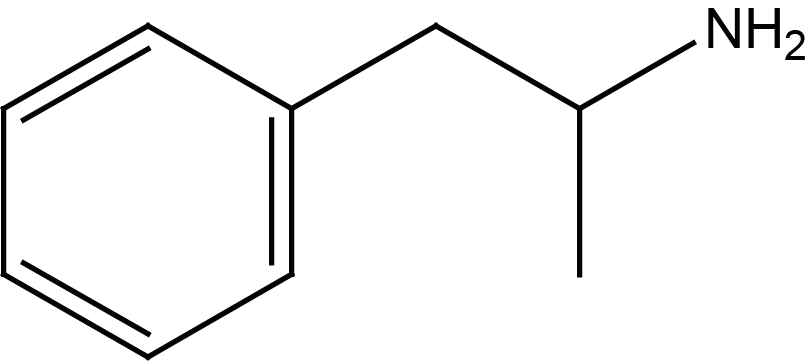}}&	& 3 conformers~\cite{Brause_CP327_43} 																&	37\,549				\\
										&																																					&	&																																			& 37\,558				\\ 
										&																																					&	&																																			& 37\,592				\\ [0.3em] 
										&																																					&	&	(H$_2$O)$_1$-cluster~\cite{Brause_CP327_43,Yao_JPCA104_6197} 				& 37\,574				\\
										&																																					&	&																																			& 37\,578				\\ 
										&																																					&	&																																			& 37\,617				\\	
\hline\hline
\end{tabular}
\end{table*}

\begin{table*}
\caption{Molecular systems and their water clusters suitable for conformer selection - \emph{continued}.}
\label{tab:systems_2}
\begin{tabular}{L{4cm}C{5cm}cC{4cm}C{3cm}}
\hline\hline
Substance								&	Structure																																	&	&																				& Electronic origin (cm$^{-1}$)	\\
\hline
\multicolumn{5}{c}{\underline{Protected amino acids}}	\\ [0.5em]
N-acetyl phenylalanine 	& 	\multirow{5}{*}{\includegraphics[width=3.839cm]{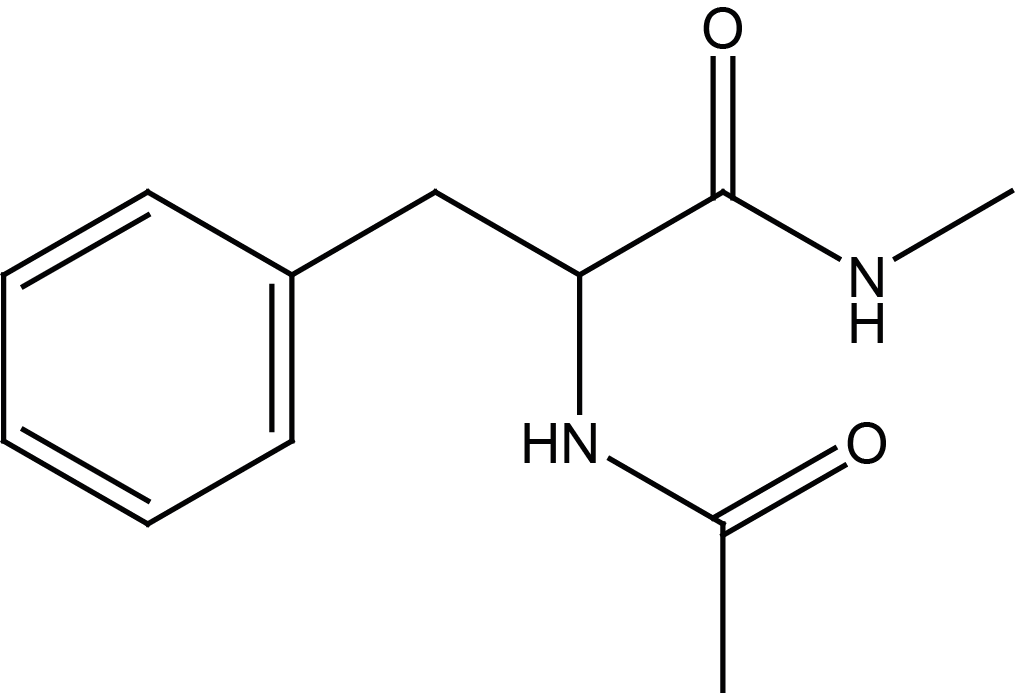}}	&	&	3 conformers~\cite{Gerhards_PCCP6_2682}	&	37\,414		 	\\ 
methyl amide						&																																						&	&																					&	37\,518			\\
												&																																						&	&																					&	37\,593			\\
												&																																						&	&																					& 						\\	
												&																																						&	&																					& 						\\	
												&																																						&	&																					& 						\\
												&																																						&	&																					& 						\\
N-acetyltryptophan 			&	 	\multirow{5}{*}{\includegraphics[width=3.984cm]{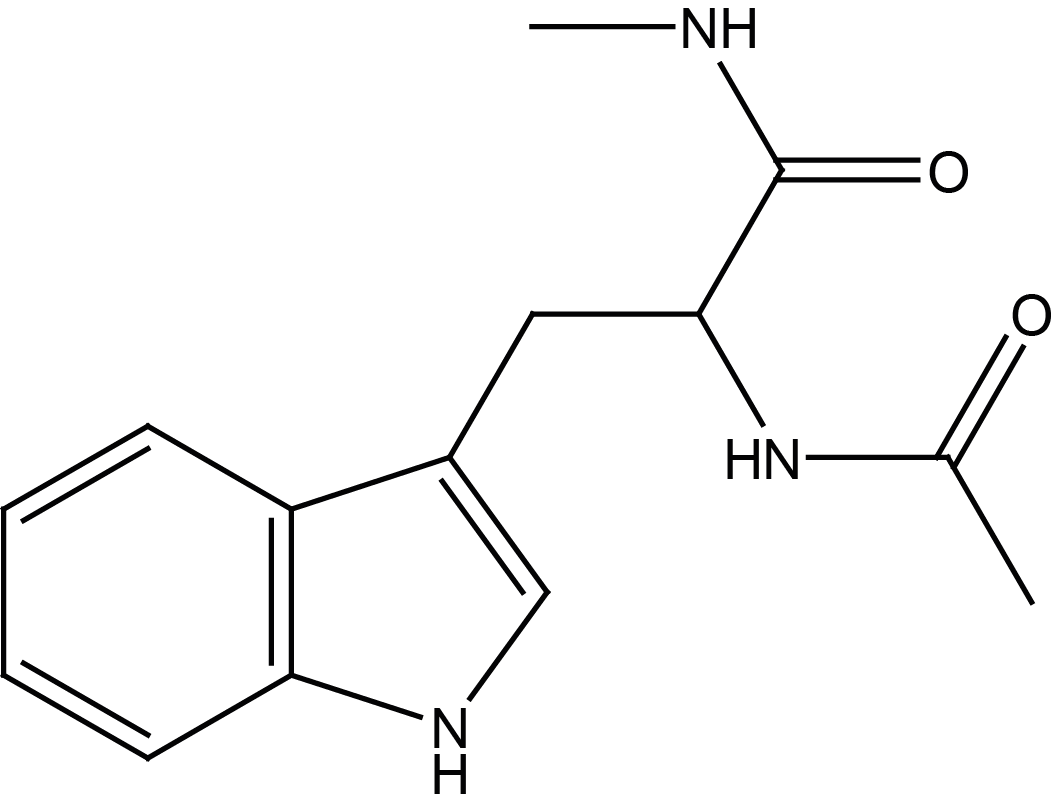}} 			&	& 3 conformers~\cite{Dian_JCP117_10688} 	&	34\,881			\\  
methyl amide						&																																						&	&																					&	34\,913			\\ 
												&																																						&	&																					&	-						\\
												&																																						&	&																					& 						\\	
												&																																						&	&																					& 						\\	
												&																																						&	&																					& 						\\
												&																																						&	&																					& 						\\
												&																																						&	&																					& 						\\
N-acetyl tryptophan  		&	 	\multirow{3}{*}{\includegraphics[width=3.996cm]{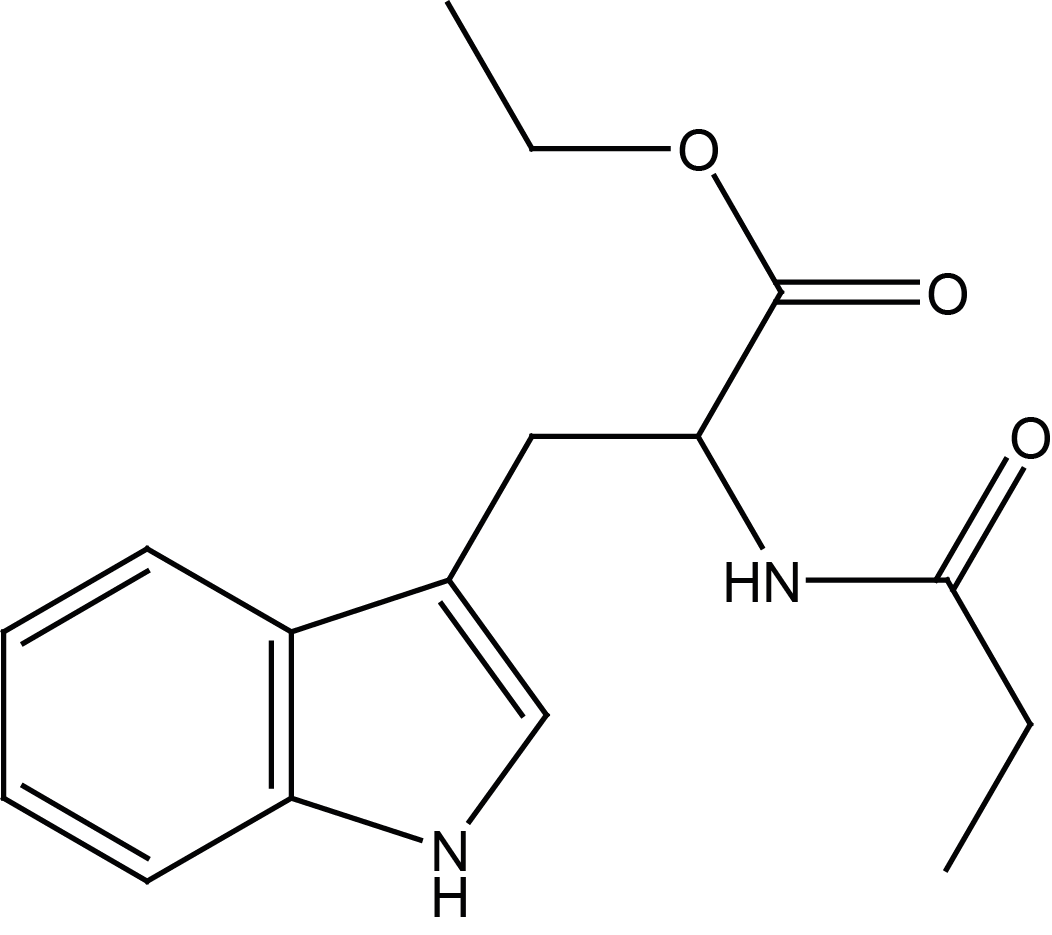}}				&	&	4 conformers~\cite{Park_JCP84_6539} 		&	34\,694			\\ 
ethyl ester							&																																						&	&																					&	34\,832			\\ 
												&																																						&	&																					&	34\,855			\\ 
												&																																						&	&																					&	34\,999			\\ 
												&																																						&	&																					& 						\\	
												&																																						&	&																					& 						\\	
												&																																						&	&																					& 						\\
												&																																						&	&																					& 						\\
												&																																						&	&																					& 						\\				
\multicolumn{5}{c}{\underline{Tautomers}}	\\ [0.5em] 
2-Hydroxypyridine 			&	\multirow{3}{*}{\includegraphics[width=4.332cm]{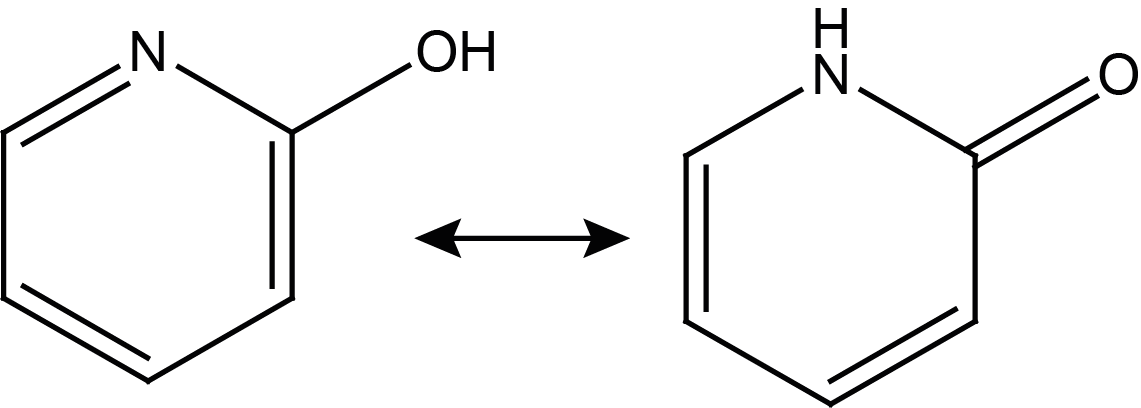}}			&	& 2 tautomers~\cite{Nimlos_JPC93_643,Held_JACS115_9708} &	29\,831	\\ 
												&																																										&	&																												& 36\,136 \\ [0.3em]
												&																																										&	&	(H$_2$O)$_1$-cluster~\cite{Nimlos_JPC93_643,Held_JACS115_9708}& 35\,468	\\ 
												&																																										&	&																						& 				\\ 
\multicolumn{5}{c}{\underline{Sugars}}	\\ [0.5em]                                                                    
Phenyl $\beta$-					&	\multirow{3}{*}{\includegraphics[width=4.584cm]{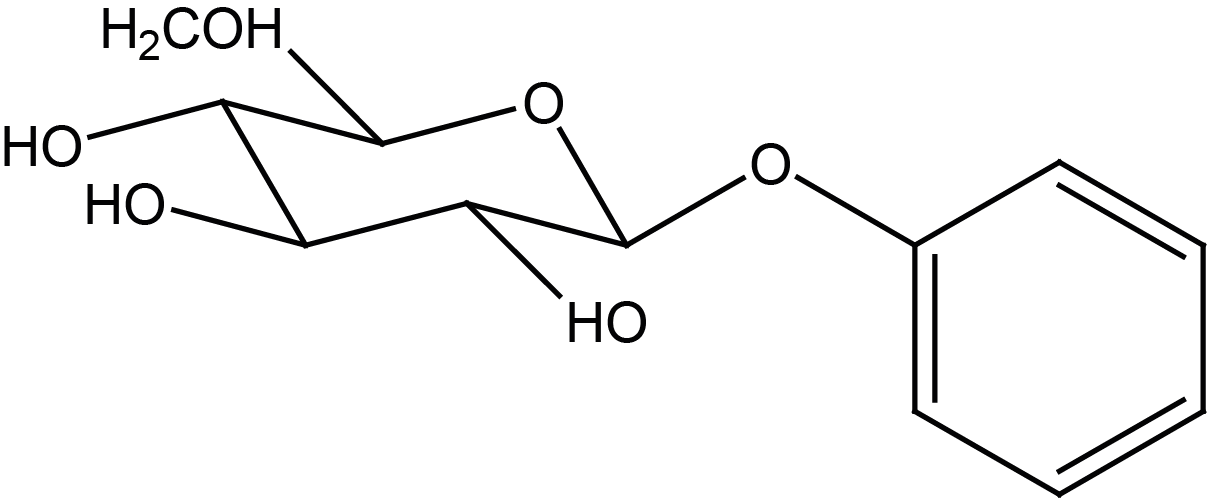}}			&	& 3 conformers~\cite{Talbot_PCCP4_3562}			& 36\,868 \\ 
D-glucopyranoside				&																																										&	&																						& 36\,903 \\ 
												&																																										&	&																						& 36\,906	\\ [0.3em]
												&																																										&	&	(H$_2$O)$_1$-cluster~\cite{Jockusch_JPCA107_10725}& 36\,767 				\\ 
												&																																										&	&																						& 36\,870	\\ [0.5em]
Phenyl $\beta$-					&	\multirow{3}{*}{\includegraphics[width=4.266cm]{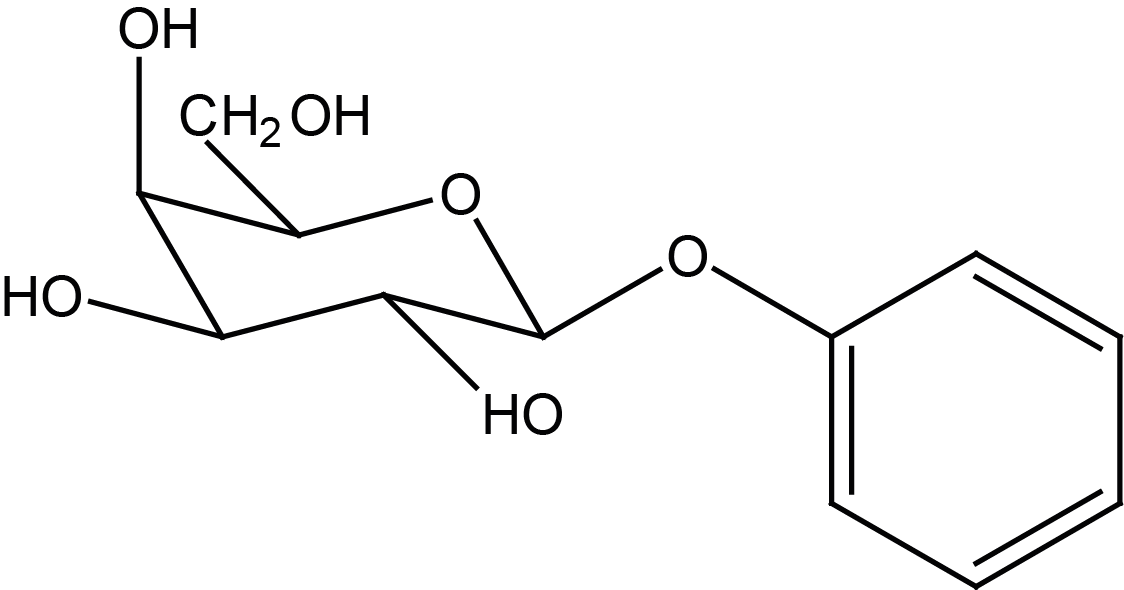}}		&	&	2 conformers~\cite{Jockusch_PCCP5_1502} 	&	36\,839 \\ 
D-galactopyranoside			&																																										&	&																						& 36\,854 \\ 													
												&																																										&	&																						& 				\\ 
												&																																										&	&																						& 				\\ 
												&																																										&	&																						& 				\\ 
												&																																										&	&																						& 				\\ 											
\multicolumn{5}{c}{\underline{Aromatic radicals}} \\ [0.5em]
$\alpha$-Propyl benzyl  &	\multirow{3}{*}{\includegraphics[width=3.294cm]{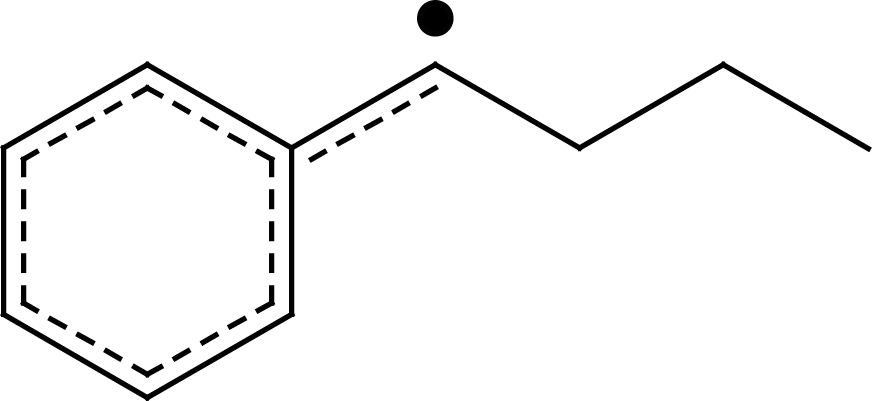}} 	&	&	2 conformers~\cite{Korn_JCP145_124314}		&	21\,922\footnote{The electronic origin is not known. This wavenumber corresponds to a strong vibrational band from the IR-UV hole-burning spectrum.}\\ 
radical									&																																										&	&																						& 21\,929	\\ 
												&																																										&	&																						& 				\\ 
												&																																										&	&																						& 				\\ 
\hline\hline
\end{tabular}
\end{table*}

\begin{table*}
\caption{Molecular systems and their water clusters suitable for conformer selection - \emph{continued}.}
\label{tab:systems_3}
\begin{tabular}{L{4cm}C{5cm}cC{4cm}C{3cm}}
\hline\hline
Substance						&	Structure																																					&	&																				& Electronic origin (cm$^{-1}$)	\\
\hline
\multicolumn{5}{c}{\underline{Aromatic hydrocarbons}}\\ [0.5em]
Propylbenzene				&	\multirow{3}{*}{\includegraphics[width=2.195cm]{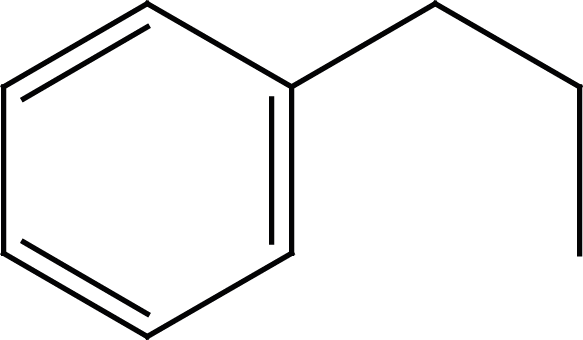}}				& &	2 conformers~\cite{Dickinson_JChemSocFaradayTrans93_1467}	&	37\,533	\\
										&																																										&	&																														& 37\,583	\\ 
										&																																										&	&																														& 				\\ 
										&																																										&	&																														& 				\\ 										
Butylbenzene 				& \multirow{3}{*}{\includegraphics[width=3.292cm]{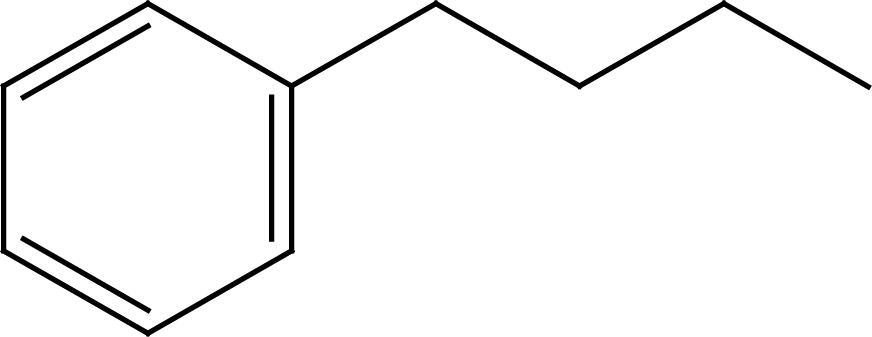}}					& &	4 conformers~\cite{Dickinson_JChemSocFaradayTrans93_1467}	&	37\,514	\\
										&																																										&	&																														& 37\,516	\\ 
										&																																										&	&																														& 37\,573	\\ 
										&																																										&	&																														& 37\,576	\\  [0.3em]
4-phenyl-1-butene 	&	\multirow{3}{*}{\includegraphics[width=3.294cm]{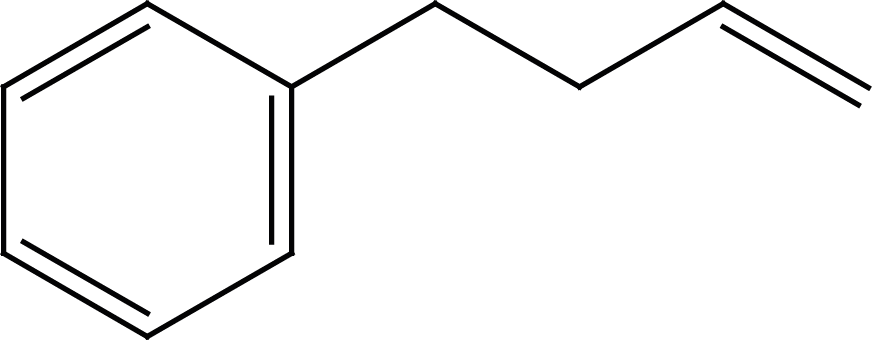}} 		& &	3 conformers~\cite{Sebree_FaradayDiscuss147_231}					&	37\,525	\\
										&																																										&	&																														& 37\,528	\\ 
										&																																										&	&																														& 37\,580	\\ [0.3em]
										&																																										&	&																														& 				\\ 
1,3-Diethylbenzene	&	\multirow{3}{*}{\includegraphics[width=3.289cm]{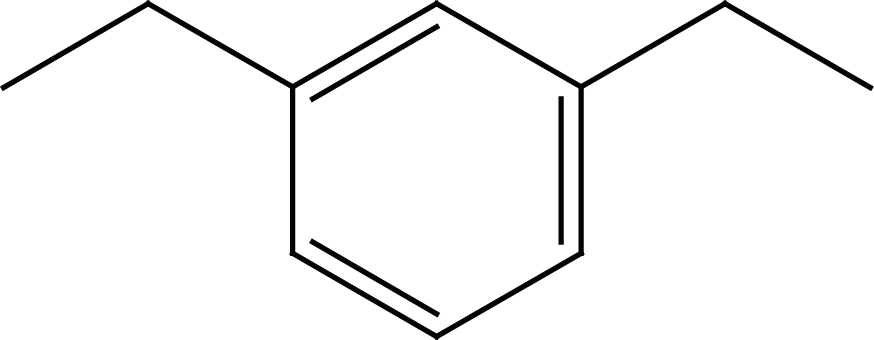}}		& & 2 conformers~\cite{Breen_JCP87_3269} 											&	37\,134	\\ 
										&																																										&	&																														& 37\,151	\\ 
										&																																										&	&																														& 				\\ 
										&																																										&	&																														& 				\\ 
Meta-Ethnylstyrene 	& \multirow{3}{*}{\includegraphics[width=2.272cm]{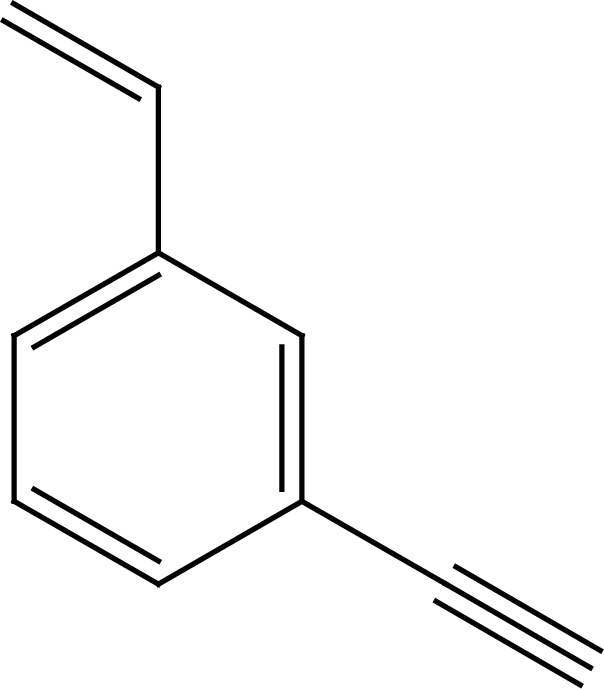}}		& & 2 conformers~\cite{Selby_JPCA109_4484}										&	32\,672	\\
										&																																										&	&																														& 32\,926	\\ 
										&																																										&	&																														& 				\\ 
										&																																										&	&																														& 				\\ 
										&																																										&	&																														& 				\\ 
										&																																										&	&																														& 				\\ 
										&																																										&	&																														& 				\\ 
\multicolumn{5}{c}{\underline{Size-selected water cluster}}\\ [0.5em]									
Phenol(H$_2$O)$_n$ 	& \multirow{3}{*}{\includegraphics[width=3.676cm]{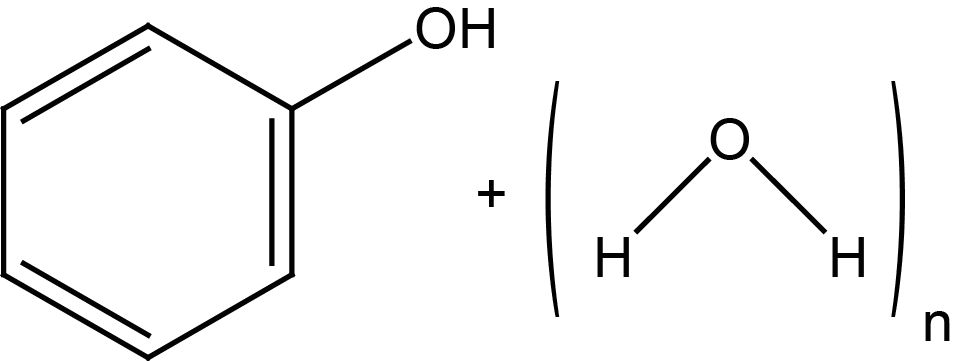}}					& &	n=1~\cite{Jacoby_JPhysChemA102_4471}											&	35\,996	\\
										&																																										&	&	n=3~\cite{Jacoby_JPhysChemA102_4471}											& 36\,258	\\ 
										&																																										&	& n=4~\cite{Jacoby_JPhysChemA102_4471}											& 36\,170	\\ 
										&																																										&	&	n=5~\cite{Jacoby_JPhysChemA102_4471}											& 36\,297	\\  	
										&																																									 	& &	n=8~\cite{Roth_CP239_1}																		& 35\,923	\\  [0.3em]								
\hline\hline
\end{tabular}
\end{table*}

\end{document}